\documentclass[onecolumn]{revtex4}
\usepackage{graphicx}
\begin{document}
\title{Bistability in a simple fluid network due to viscosity contrast}
\date{\today}
\author{John B. Geddes}
\email[Address correspondence to:]{john.geddes@olin.edu}
\affiliation{Franklin W. Olin College of Engineering, Needham MA}
\affiliation{Institute for Complex Systems and Mathematical Biology, University of Aberdeen, Scotland}
\author{Brian D. Storey}
\affiliation{Franklin W. Olin College of Engineering, Needham MA}
\author{David Gardner}
\affiliation{Franklin W. Olin College of Engineering, Needham MA}
\author{Russell T. Carr}
\affiliation{Department of Chemical Engineering, University of New Hampshire, Durham NH}
\begin{abstract}
We study the existence of multiple equilibrium states in a simple fluid network using Newtonian fluids and laminar flow. We demonstrate theoretically the presence of hysteresis and bistability, and we confirm these predictions in an experiment using two miscible fluids of different viscosity---sucrose solution and water. Possible applications include bloodflow, microfluidics, and other network flows governed by similar principles.
\end{abstract}
\maketitle

\section{Introduction}

Bistability and oscillations are ubiquitous in natural systems, and occur on a variety of scales from the micro (e.g. microfluidics) to the macro (e.g. magma flow). While the physics often differs, the necessary ingredients are a source of nonlinearity and a feedback mechanism. For example, the flow of droplets through microfluidic networks demonstrates bistability and oscillations, even when the network boundary conditions are fixed \cite{Jousse:2006p7,Schindler:2008p4582}. The source of nonlinearity and feedback in this case is the hydrodynamic resistance due to the presence of droplets in various parts of the network \cite{Belloul:2009p4409}. On the macro-scale, bistability and oscillations have been predicted in models of magma flow due to either temperature-dependent viscosity \cite{Helfrich:1995p5597} (magma becomes more viscous as it cools) or volatile-dependent viscosity \cite{Wylie:1999p3602} (magma becomes more viscous as it loses water content). In these cases, the existence of multiple solutions on the equilibrium pressure-flow curve leads to bistability and spontaneous oscillations.

Bistability and oscillations in fluid networks are also key ingredients for fluidic logic devices. In the last ten years, multiple researchers have demonstrated that microfluidic memory, logic, and control devices can be constructed using a single-phase fluid with elastic properties \cite{Groisman:2003p4579,Groisman:2004p4580} or using droplets or bubbles which form spontaneously when two immiscible fluids merge \cite{Fuerstman:2007p10,Prakash:2007p4, Joanicot:2005p4472}.
These recent studies are reminiscent of classic fluidic circuits which were quite common in the 1960s \cite{Foster1970}. In classic fluidics the nonlinearity needed to build devices such as logic gates and flip-flops comes from inertial flow effects, which is not present in microfluidic systems.

Another important fluid network  which may exhibit similar behavior is the microvascular network.
Modeling and simulation of microvascular bloodflow involves a number of non-linear rheological effects that are absent in larger vessels (see \cite{Popel:2005aa} for a good review). The Fahraeus-Lindqvist effect refers to the viscosity of blood as a function of both hematocrit (red blood cell volume concentration) and vessel diameter. The apparent reduction in viscosity as diameter decreases is usually attributed to the existence of a lubricating layer of plasma along the edge of the vessel. This lubricating layer is also responsible for the plasma skimming effect which describes the flow-dependent separation of blood at diverging bifurcations.

Over the last ten years, we (JBG and RTC) have focused our attention on analyzing the flow of blood through small networks of microvessels. We have demonstrated that, in a simple network consisting of a single inlet, a loop, and a single outlet multiple equilibria could exist and that steady state solutions could become unstable via a Hopf bifurcation to a limit cycle oscillation \cite{Carr:2005p42,Geddes:2007aa}. We recently turned our attention to the three-node network (Figure \ref{fig:network}) which consists of two inlets, a loop, and one outlet. The flow in vessel C can be clockwise or counter-clockwise depending on the equilibrium state. Using models of the Fahraeus-Lindqvist effect and the plasma skimming effect appropriate for {\em in vivo} bloodflow \cite{Pries:1990p1493}, we have demonstrated that realistic network geometries exist which can support flow in either direction in vessel C, i.e. bistability \cite{Gardner:2009p4875}.

In this paper we explore bistability in a system where essentially all the complexities which exist in many of the above studies have been removed. We consider the low Reynolds number laminar flow of two ordinary miscible Newtonian fluids  of different viscosity in the simple network shown in Figure \ref{fig:network}. Remarkably, we find theoretically and experimentally that bistability can exist in this system which, at first glance, seems linear. However, when ordinary fluids mix, the viscosity of the resulting mixture depends (nonlinearly) on the relative volume fraction of each fluid and this nonlinearity is sufficient to introduce bistability in the network flow.  After reviewing the model and the theoretical predictions we discuss the experimental set-up and present results which confirm our analysis. We end with a discussion of the implications of these results.

\section{Model}

\begin{figure}
\includegraphics[width=6cm]{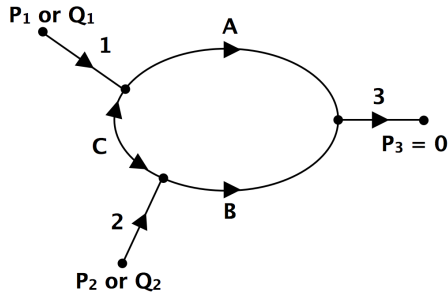}
\caption{Water enters through inlet 1 and a water-sucrose mixture enters through inlet 2. The network can either be flow-driven or pressure-driven. The flow in branch C can be either clockwise or counter-clockwise depending on the equilibrium flow distribution. The outlet is held at a fixed pressure.}
\label{fig:network}
\end{figure}

We explore flow in the simple three-node network shown in Figure \ref{fig:network}. Two miscible fluids of different viscosities are injected in inlet 1 and 2 at either fixed pressure or flow. The outlet is held at a fixed pressure. The network model assumes laminar, fully developed flow in each of the branches with well-mixed fluid after junctions where the two fluids meet. Similar results can be obtained even if the fluids do not mix after junctions, and we consider that case later in the paper.
Without loss of generality, we assume that the more viscous fluid always occupies inlet 2.

We assume that the fluid obeys Poiseuille's law such that the pressure drop $\Delta P$ is proportional to the volumetric flow $Q$,
\begin{equation}
\Delta  P = R Q = \frac{128 \mu L}{\pi D^4 } Q,
\label{eq:pois}
\end{equation}
where $R$ is the hydraulic resistance, $D$ is the diameter of the tube, $L$ is the length of the tube, and $\mu$ is the fluid viscosity. 
For convenience we will refer to the nominal resistance as the part of the tube resistance which is only dependent
on the geometry and not the fluid, i.e.  $128 L/\pi D^4$. 
The total pressure drop around the loop must be zero,
\begin{equation}
\Delta P_A  = \Delta P_B + \Delta P_C,
\label{eq:press}
\end{equation}
where $\Delta P_A$ and $\Delta P_B$ are positive with the flow toward the exit. While the direction of the flow in branch C may be in either direction, we define $\Delta P_C$ to be positive when the flow is counter-clockwise, i.e. from inlet 1 to inlet 2. Since the liquid is incompressible the total flow rate into any junction must also equal the flow rate out, namely $Q_A =Q_1 - Q_C$ and $Q_B  = Q_2+Q_C$. When the fluids mix at a node,  their volumes are additive such that we can compute the volume fraction of fluid 2, $\phi$, in the resulting mixture. Using this volume fraction to characterize the mixture composition means that $\phi_1 =0$ and $\phi_2=1$. When $Q_C$ is positive (from inlet 1 to inlet 2) the volume fraction in each tube is
\begin{eqnarray}
 \phi_A = 0, \quad \phi_B = \frac{Q_2}{Q_C + Q_2}, \quad \phi_C = 0.
\label{eq:phib}
\end{eqnarray}
When $Q_C$ is negative (from inlet 2 to inlet 1) the volume fraction in each tube is
\begin{eqnarray}
\phi_A = \frac{Q_C }{Q_C - Q_1}, \quad \phi_B = 1, \quad \phi_C = 1.
\label{eq:phia}
\end{eqnarray}
The viscosity of the mixture can be captured using a modified Arrenhius law,
\begin{equation}
\mathrm{log}(\mu) = (1-N) \mathrm{log}(\mu_1) + N \mathrm{log}(\mu_2) + N (1-N) d,
\label{eq:visc}
\end{equation}
where $N$ is the mole fraction of fluid 2 and d is a fit parameter. The first two terms on the right hand side are the classic Arrenhius law and third term is an empirical correction factor \cite{Grunberg:1949p4869}. We use this relationship to fit experimental data for real mixtures \cite{RLide:2007p4708}.

\begin{figure}
a) \includegraphics[width=8.5cm]{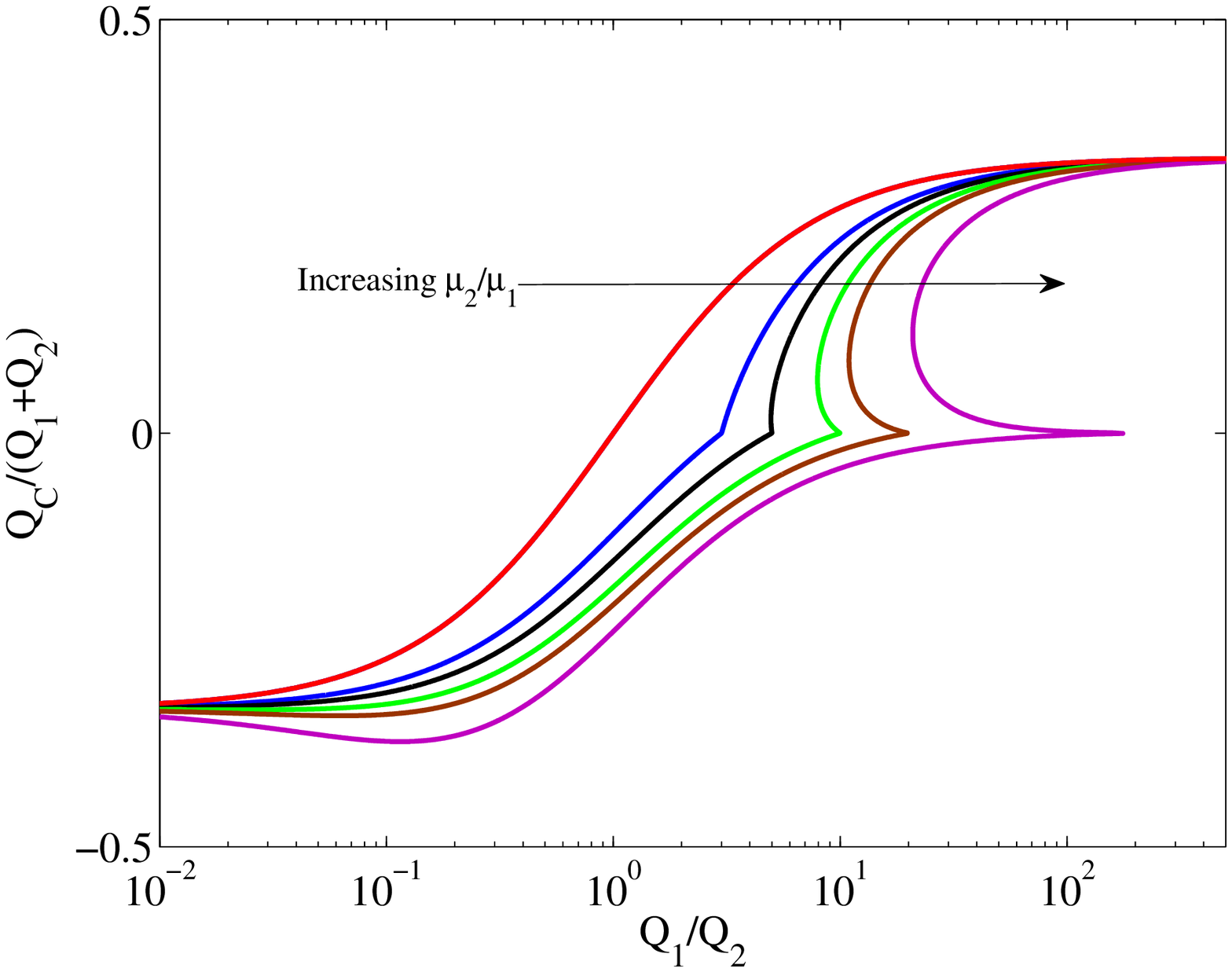}
b) \includegraphics[width=8.5cm]{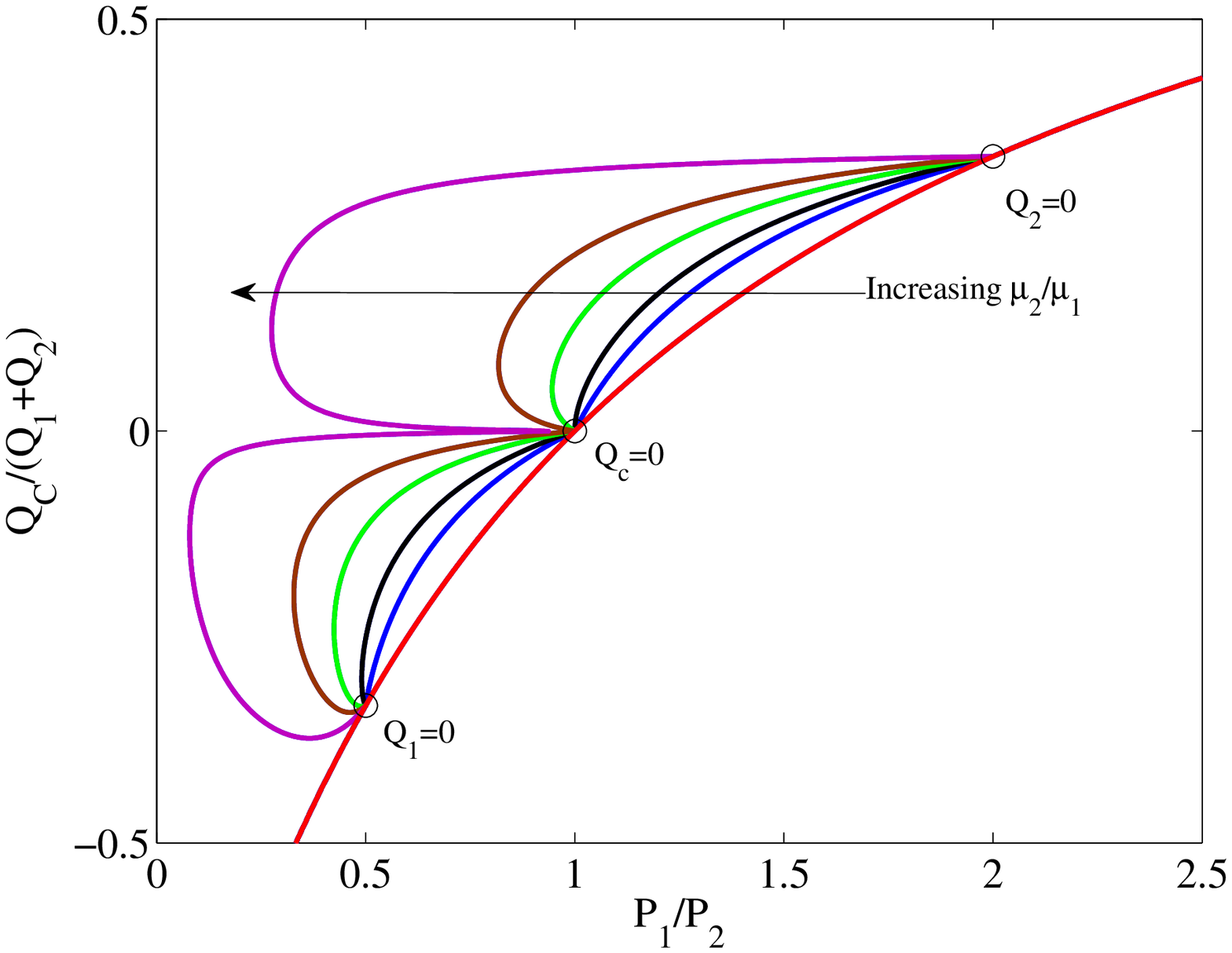}
\caption{ (Color online) a) Normalized flow rate in branch C ($Q_C/(Q_1+Q_2)$) as a function of the inlet flow ratio $Q_1/Q_2$ for different viscosity ratios. Here we have taken the nominal resistance of every branch in the network to be equal, we assume a perfect Arrenhius law for viscosity, and we assume that the fluids are otherwise identical except for their viscosity. Changing the nominal resistance of each branch leads to qualitatively similar results. Results are shown for viscosity ratios of 1, 3, 5, 10, 20, and 200. Increasing the viscosity ratio increases the width of the region of bistability.
b) Normalized flow rate in branch C as a function of the inlet pressure ratio $P_1/P_2$ for the same viscosity ratios and conditions
as in the flow case.
The nominal resistance of the inlets and the outlet is set equal to the resistance in all other branches. The values of
$P_1/P_2$ where the flow in branches C, 1, and 2 go to zero are denoted.}
\label{fig:hysteresis}
\end{figure}

Using Eq. \ref{eq:press} and the conservation of flow, the flow in C can be expressed in terms of the inlet flow in 1 and 2,
\begin{equation}
Q_C  = \frac{Q_1 R_A - Q_2 R_{B}}{ R_A  + R_B + R_C}.
\label{eq:qc}
\end{equation}
In this way, Eqs. (\ref{eq:pois})-(\ref{eq:qc}) define a nonlinear, algebraic equation for the flow in branch C which is parameterized by the inlet flows, the viscosity ratio between the fluids, and the length and diameter of the tubes. It is possible to find network parameters, i.e. tube lengths, diameters, and inlet fluid viscosities such that the equilibrium flow in branch C demonstrates hysteresis or bistability as the inlet flow rates are varied. In Figure \ref{fig:hysteresis} (a) we show examples of the normalized flow rate in branch C as a function of the ratio of the inlet flow
rates for different viscosity ratios. For low and high inlet flow ratio values the normalized flow in branch C is unique, but there exists a set of inlet flow ratios which leads to multiple equilibrium flow values in branch C. As the viscosity ratio of the two fluids increases, the window where the system exhibits bistability widens. The point where the flow in branch C goes to zero is simply $Q_1/Q_2 = R_B/R_A$. For a symmetric network where the nominal resistance in A and B are equal, the flow in branch C goes to zero when $Q_1/Q_2 = \mu_2/\mu_1$. This is the case in Figure \ref{fig:hysteresis}a.

We may also drive the network by applying fixed pressure at inlet 1 and 2, as opposed to a fixed flow.  Under pressure-driven conditions, Eqs. (\ref{eq:pois}-\ref{eq:qc}) are supplemented with equations for the inlet pressures (relative to the exit pressure),
\begin{eqnarray}
\frac{P_{1}}{P_2}  =\frac{ (Q_1 + Q_2) R_3 + (Q_1 - Q_C) R_A + Q_1 R_1}
{(Q_1 + Q_2) R_3 + (Q_2 + Q_C) R_B + Q_2 R_2}.
\label{eq:P}
\end{eqnarray}
Under pressure-driven conditions the resistance of the inlet and outlet tubes enters the formulation. However, if the equilibrium flow in branch C demonstrates bistability under flow-driven conditions, then bistability will be observed under pressure-driven conditions about the point of zero flow in branch C. In Figure \ref{fig:hysteresis} (b) we show examples of the equilibrium curve for the flow in branch C as a function of the corresponding inlet pressure ratio.

Under pressure-driven conditions, there exists a region of bistability around the point where the flow in branch C goes to zero. We also find a second region of bistability which occurs around the point of zero flow in inlet branch 1. This second region of bistability corresponds  to back-flow into the supply reservoir 1, the low viscosity fluid. The points where the flow in branches C, 1, and 2 go to zero are easily calculated. For a perfectly symmetric network, where all the nominal resistances are equal, these points are defined by
\begin{equation}
Q_c=0 ~~ \mathrm{when}~~ \frac{P_1}{P_2}=1; ~~
Q_1=0 ~~ \mathrm{when}~~ \frac{P_1}{P_2}=\frac{1}{2}; ~~
Q_2=0 ~~ \mathrm{when}~~ \frac{P_1}{P_2}=2.
\end{equation}
This result is easily generalized to any arbitrary network. We are assuming a symmetric network here only for simplicity.

In Figure  \ref{fig:hysteresis} (b) we show that when  $P_1/P_2 > 2$ the flow is into the network through branch 1 and out of the network through branches 2 and 3, i.e. there is backflow into reservoir 2. In this domain the dimensionless flow, $Q_C/(Q_1+Q_2)$, has a unique value for a given pressure ratio $P_1/P_2$, which is the same for all viscosity ratios. When $1 < P_1/P_2 < 2$, the flow is into the network from branches 1 and 2 and the flow in C is also always positive. In this domain there is a unique value of $Q_C$ for a given pressure ratio, but its value depends upon the viscosity contrast. When $0.5 < P_1/P_2 < 1$  the flow is still into the network through branches 1 and 2. However a region of bistability may occur in this domain in which the flow in branch C can be either clockwise or counter-clockwise. The width of this region of bistability widens as the viscosity ratio increases. When  $P_1/P_2 < 0.5$, the flow is into the network through branch 2 and a second region of bistability exists with respect to the flow in branch 1. At sufficiently low values of $P_1/P_2$ the flow is always into the network through branch 2 and out of the network through branches 1 and 3. For a viscosity ratio of 200 (left blue curve), the point where the flow must always be going out of the network at branch 1 occurs at $P_1/P_2=0.08$. It is interesting to note that at sufficiently high viscosity ratios, there may exist 5 equilibrium flow conditions for a given pressure ratio, 3 of which are stable and 2 of which are unstable. It was surprising to us to find such rich behavior in a simple network under laminar flow with Newtonian fluids. For the remainder of the paper, we concentrate on the bistability region about the point $Q_C=0$.

\section{Experiments}
In order to test our prediction of bistability, we built a simple and inexpensive experimental flow network. The flow network was constructed with common 1/16 inch inner diameter clear plastic tubing. The connections in the network are made with barbed T fittings. In order to have well-mixed flow in the tubes, we place static inline flow mixers right after the T-junctions in branches A and B. Without the mixers, the two fluids remain stratified (they were not density matched) which changes the switching points but does not destroy the hysteresis.

Inlet tubes 1 and 2 are connected to a pressure source, which is simply a large graduated cylinder filled to a set level. The graduated cylinders act as our flow reservoirs and the pressure of the reservoirs are known from hydrostatics. The reservoir levels change slowly as the network drains, but this draining can be offset by slowly pumping new fluid into the reservoirs. Inlet 1 always contains pure  water while inlet 2 consists of an aqueous sucrose solution of varying concentration. Food coloring is added so that we can observe the direction of flow in branch C as well as confirm mixing. The relative lengths of the tubes were $L_A=L_B$, $L_C=\frac{6}{45}L_B$, $L_1=L_2=L_3=\frac{4}{15}L_B$ and we chose $L_B = 0.75$m for convenience. The mixers were tested experimentally and their equivalent resistance was approximately that of $0.375$m of additional tubing. Before each experiment, we measured the viscosity and density of the sucrose mixture in reservoir 2 to confirm that it matches known data \cite{RLide:2007p4708}.

\begin{figure}
a) \includegraphics[width=8cm]{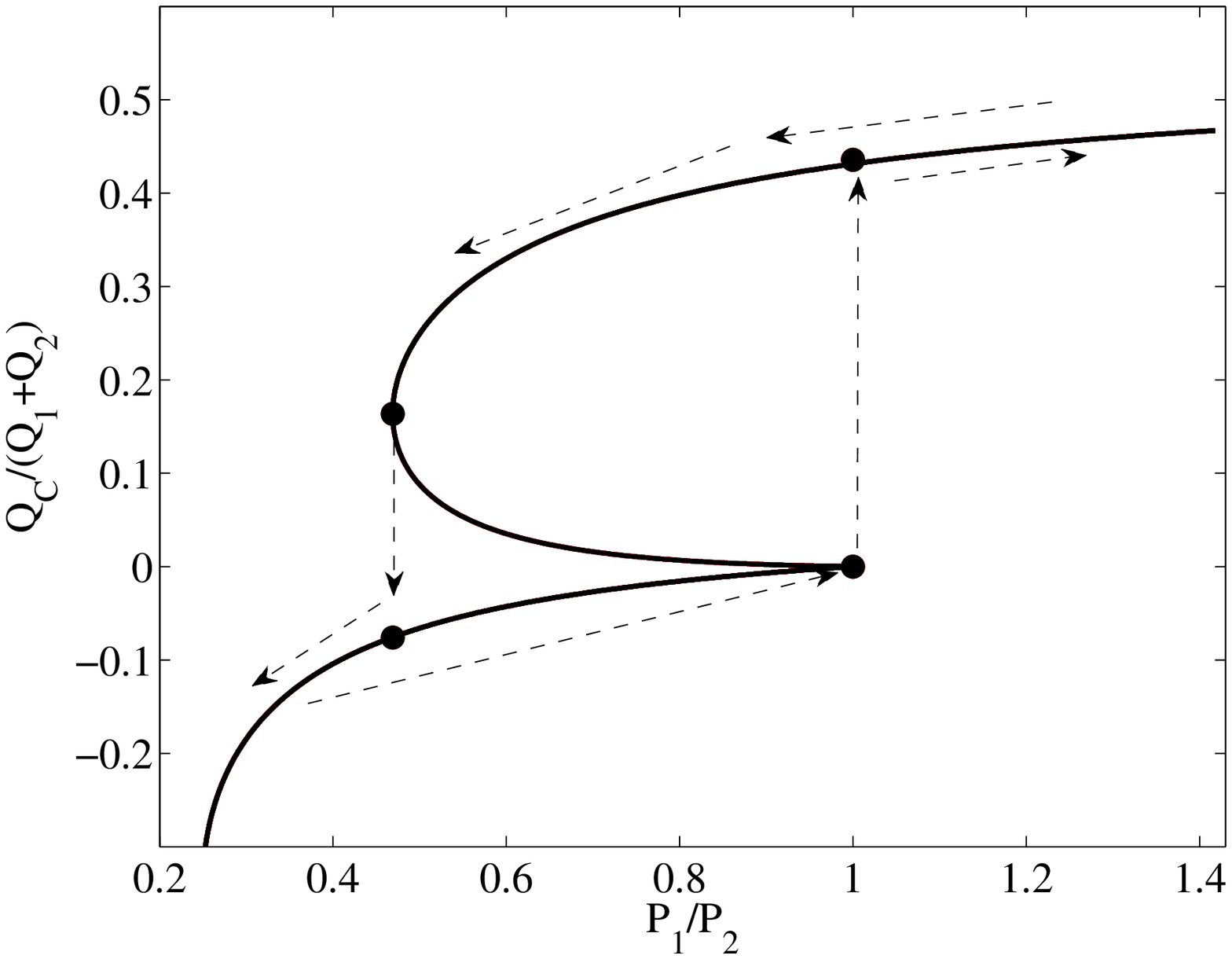}
b) \includegraphics[width=8cm]{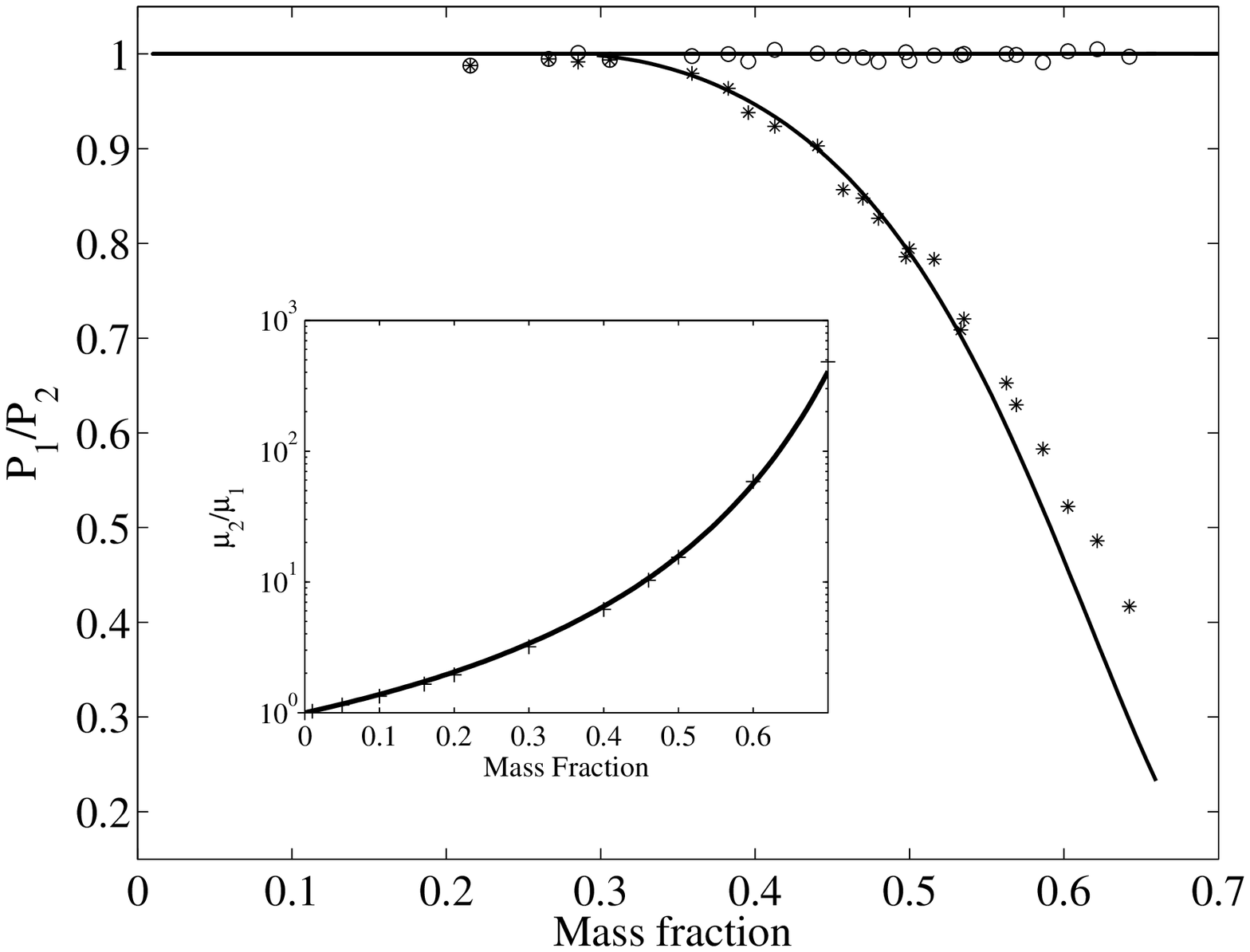}
\caption{(a) Sample equilibrium curve corresponding to the specific experimental network and a mass fraction of 60\% sucrose in
inlet 2. The hysteresis loop (dashed line) and switching points (solid circles) are indicated. (b) Switching pressure ratios versus mass fraction of sucrose in inlet 2 as determined experimentally (points) and as predicted by the model (lines). Below a concentration of approximately 30\% no bistable behavior is expected nor observed. The inset shows the relationship between mass fraction of sucrose
in the solution and its viscosity relative to water.
The points on the inset are experimental data from \cite{RLide:2007p4708} and the solid curve is a fit using the
modified Arrenhius law. }
\label{fig:bistable_exp}
\end{figure}

To observe bistability, the level of the sucrose solution in reservoir 2 is maintained at a constant level. The level of the water in reservoir 1 is initially high so that the fluid in branch C is water, $Q_c>0$ and flowing from 1 to 2. The level of the water in reservoir 1 is allowed to drop slowly as the network drains the reservoir, thus decreasing the pressure at inlet 1. The rate at which the level of reservoir 1 falls is slow compared to the time for the network to come to equilibrium. At some point we visually observe the flow in C to
become negative and start flowing from 2 to 1. We record the water level in reservoir 1 at this point. The level of reservoir 1 is then raised by slowly pumping water in such that the bistable region can be confirmed. We find the flow in C remains negative
for a sufficiently long time. As the level of reservoir 1 rises, the flow eventually switches direction back to $Q_c>0$, and we record the water level in reservoir 1 when this occurs. An example of the experimental path is shown in Figure \ref{fig:bistable_exp}a; our experiment records the two switching points. Thus, in this work we concentrate only on  the bistability region around the point $Q_c=0$.

We repeat the above experiment for a number of initial sucrose concentrations (viscosities) in reservoir 2. A comparison of the experimental data compared to the model prediction is shown in Figure \ref{fig:bistable_exp}b. We see good agreement between the model and experiment. For small sucrose concentrations there is no bistable window. As the sucrose concentration (viscosity) exceeds a critical value a bistable window opens up and widens as the sucrose concentration is further increased. The threshold sucrose concentration depends on the specific geometry of the network and is approximately 30\% mass fraction for our network, corresponding to $\mu/\mu_{H_2O} = 3$.

The existence of bistability depends on the network geometry and the viscosity of the inlet fluids. It is possible to analytically determine whether a given network and fluid combination can exhibit bistable behavior.
The point where the flow in C goes to zero is determined from Eq. \ref{eq:qc},  $Q_1 R_A = Q_2 R_B$. If the
slope of the
equilibrium flow-pressure curve switches direction at this point
then a bistable window will exist (see Figure \ref{fig:bistable_exp}a).
The analysis (details can be found in the appendix) yields a simple criteria for the existence of bistability,
\begin{equation}
\frac{\mu_2}{\mu_1} \left( \ln\left( \frac{\mu_2}{\mu_1} \right) \frac{\rho_1 M_2}{\rho_2 M_1} - 1\right)  >
\frac{L_A D_B^4}{L_B D_A^4}+\frac{L_C D_B^4}{L_B D_C^4},
\label{eq:existence}
\end{equation}
where $\rho_i$ is the mass density of fluid $i$ and $M_i$ is the molar mass of fluid $i$. The parameters on the left hand side depend only upon the fluids, while the parameters on the right hand side depend only upon the network geometry. If we apply this criteria to our experimental network, we find that $\frac{L_A D_B^4}{L_B D_A^4}+\frac{L_C D_B^4}{L_B D_C^4}=1.13$. The corresponding minimum mass fraction of sucrose to water in inlet 2 required to observe bistability in this particular network is 29 \% which agrees well with our experimental data. In addition, Eq. \ref{eq:existence} gives a lower bound on the viscosity contrast required to observe bistability in any network obtained by making the left-hand side positive and using the corresponding experimental values for the fluids. For a water-sucrose solution this value is approximately 2.3.

\section{Generalization}
Thus far we have presented the prediction of bistability under the special condition that the fluids are well mixed after any junction. We presented this case first as it is the most straightforward, simplest, and surprising
 manifestation of a more general behavior. The analysis which led to Eq. \ref{eq:existence} can be generalized for any effective resistance of the tube after two fluids merge. The more general criterion for bistability is (see Appendix),
\begin{equation}
\frac{\mu_2}{\mu_1} \left(  \left.\frac{d \ln(\mu)}{d \phi}\right|_{\phi=1} - 1\right)  >
\frac{L_A D_B^4}{L_B D_A^4}+\frac{L_C D_B^4}{L_B D_C^4},
\label{eq:existence_general}
\end{equation}
where the viscosity of the mixture $\mu$ is expressed in terms of the volume fraction $\phi$.
There is nothing in this expression that is unique to an Arrenhius mixture law for the system to exhibit bistability. Since the right hand side of Eq. \ref{eq:existence_general} must always be positive, then bistability is possible in general as long as $\left. d \ln(\mu) / d \phi \right|_{\phi=1} >1$. This result does reveal that bistability is not possible for a mixture with viscosity linearly dependent on volume fraction, e.g. weak suspensions.

The most practical extension of our experiment is to remove the mixers from the system. The mixers were introduced only to create a model experimental system to confirm the simple theory.
If the fluid mixers are removed from the junctions we observed that the flow is stratified due to the large density difference between the two fluids. Such a stratified, unmixed flow would be more representative of laminar, low Reynolds number flows, resulting from bringing two streams together, especially in microfluidic applications. The fully mixed case might only be expected in practice if the diameters were extremely small and branches sufficiently long such that molecular diffusion can mix the streams. However, on the scale of many microfluidic networks diffusion times are often relatively slow.

To make predictions about  stratified flow, we  need to know the effective mixture viscosity  
 to replace the Arrenhius law of Eq. \ref{eq:visc}. 
 In the stratified flow case, the effective viscosity is no longer a property of the fluid mixture, but 
 becomes a property of the flow. We therefore must solve for the velocity field in the tube to calculate the 
 overall hydraulic resistance. 
 
\begin{figure}
a) \includegraphics[width=8cm]{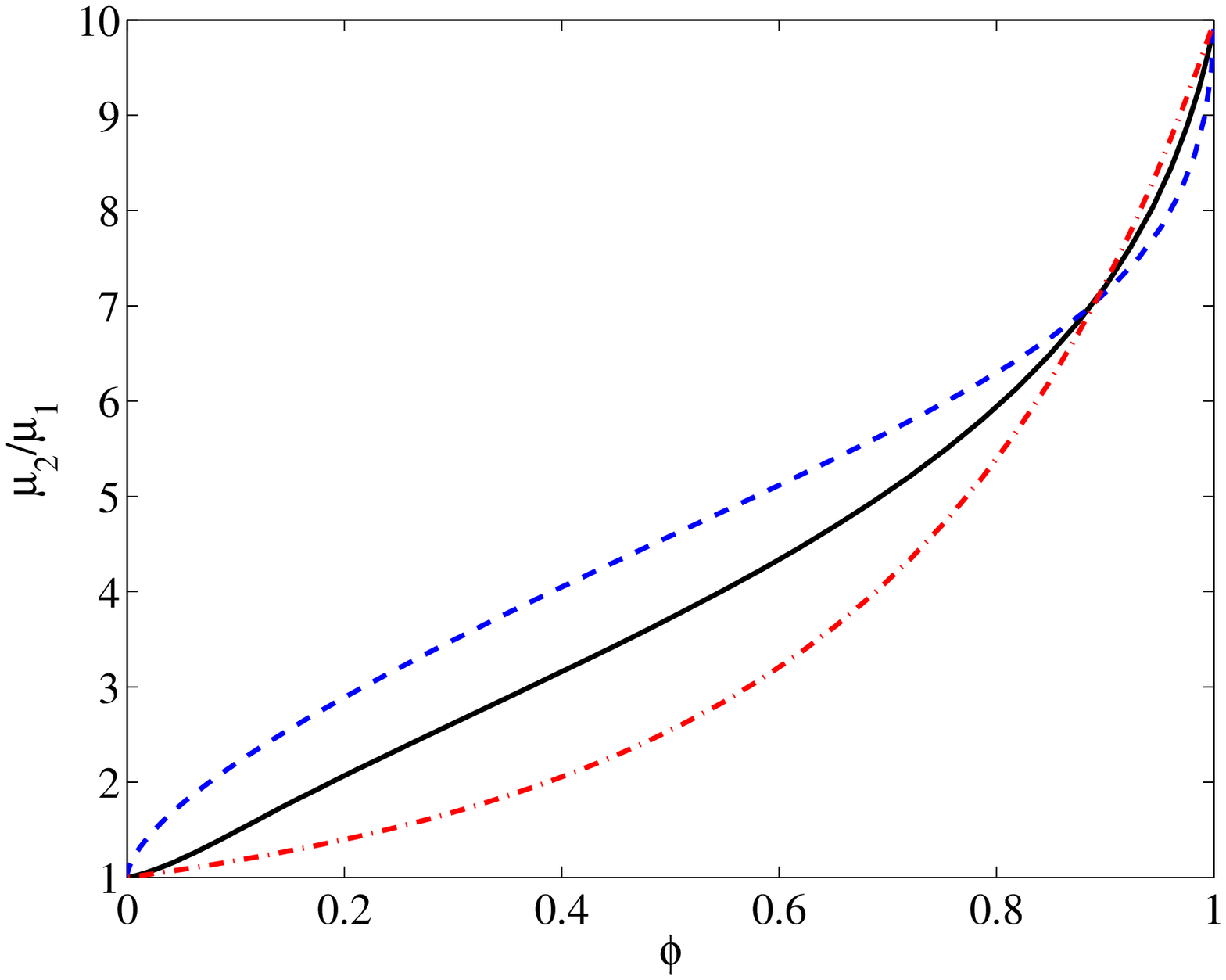}
b) \includegraphics[width=8cm]{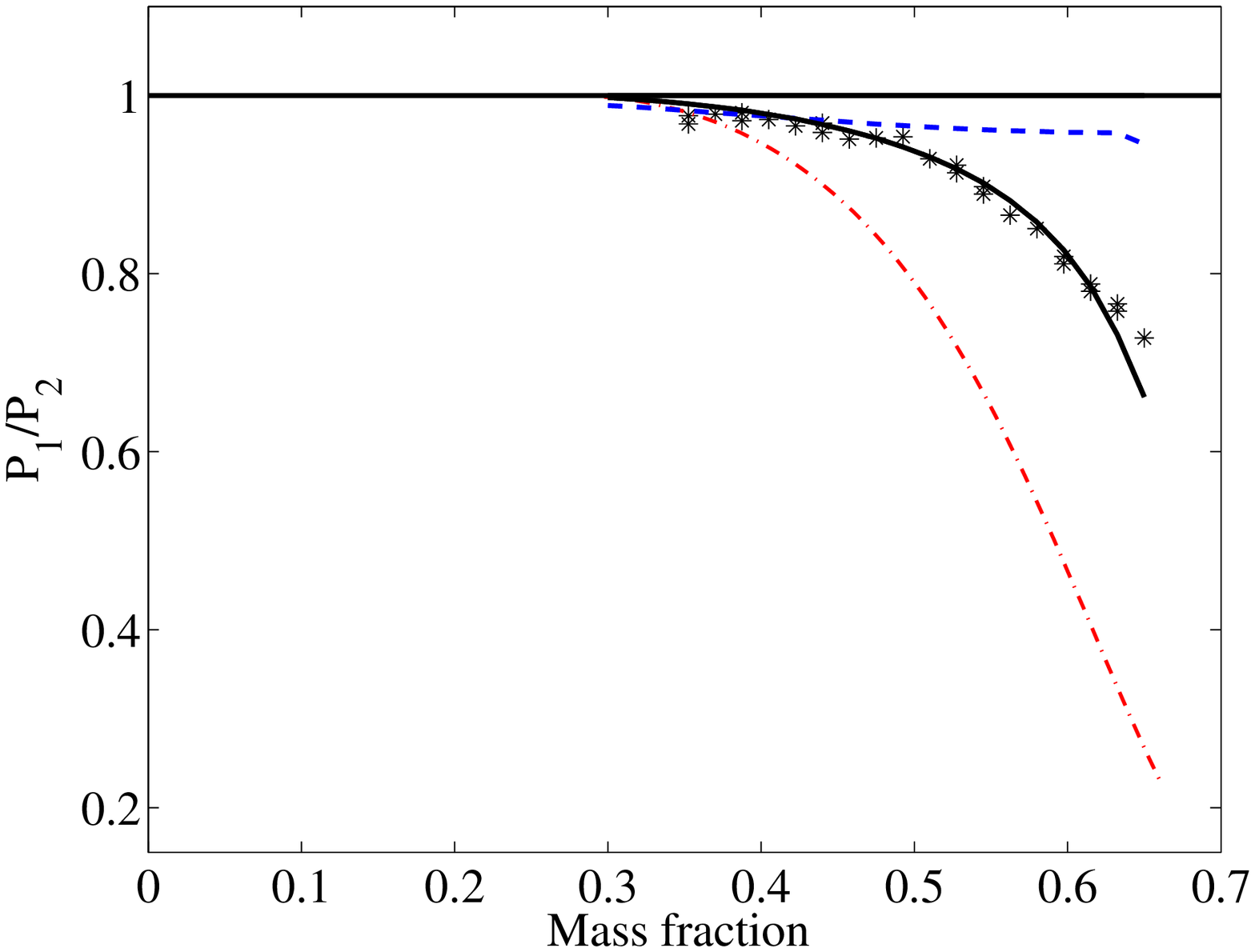}
\caption{ (Color online) a) An example of the effective viscosity $\mu/\mu_1$ of two merged fluids assuming perfect mixing (red dashed-dotted curve), perfect stratification (blue dashed curve), and some diffusion (black solid curve) as a function of volume fraction. In the diffusion curve we allowed the perfectly stratified interface to diffuse for $t=0.05$, where time is scaled by square of the tube diameter divided by the diffusivity. Here we plot the curves for a viscosity ratio of $\mu_2/\mu_1 = 10 $ only as an example.
Other viscosity ratios look qualitatively similar.
(b) Switching pressure ratio versus mass fraction of sucrose in inlet 2 as determined experimentally (points) and as predicted by the model (lines) for different degrees of stratification.
The model lines are shown for two immiscible fluids (blue dashed line), complete mixing (red dotted curve) and
some diffusion corresponding to a time of $t=0.05$ (black solid curve).
 The experiment is identical to the earlier one except the mixers have been removed. Here we present data for one switching point for clarity. The switching point at $P_1/P_2 = 1$ is the trivial point where $Q_C=0$ and the system behaves the same whether stratified or mixed.}
\label{fig:bistable_exp_strat}
\end{figure}

A simple model  defines a horizontal interface at an arbitrary location in a circular tube---fluid 1 and fluid 2 occupy the regions above and below this interface. We assume the flow is steady and
fully developed in the axial direction, $x$,  and 
solve for the Navier-Stokes equations for the 
axial velocity field, $u$, across the tube's cross section for a unit pressure drop.
The full Navier-Stokes equations become quite simplified in the steady, fully developed case; 
\begin{equation}
\nabla \cdot ( \mu \nabla {u} )= \frac{dP}{dx} = -1.
\label{eq:strat}
\end{equation}
The vector operators above are two dimensional and only apply over a cross section of tubing. 
Numerically, we can easily solve  Eq. \ref{eq:strat} using a commercial finite element package, Comsol Multiphysics.
After solving Eq. \ref{eq:strat}, the flow rates of the two fluids, $Q_1$ and $Q_2$,  
are determined from integration of the velocity field.  
The overall hydraulic resistance, and thus the effective viscosity,  is  $R=1/(Q_1+Q_2)$. 
We then  move the location of the horizontal interface between 
the two fluids from the bottom of the tube to the top which allows us to  
 calculate the resistance as a function of the volume fraction as shown by the 
 blue dashed curve in Figure \ref{fig:bistable_exp_strat}a.  This
  simple calculation reproduces the resistance law calculated by Gemmel and Epstein  in 1962 \cite{Gemmell:1962p5510}. 
  The function shown in \ref{fig:bistable_exp_strat}a then replaces the Arrenhius law of Eq. \ref{eq:visc}.
  It is interesting to note that in the case of miscible fluids this model is relatively sensitive to even small amounts of molecular diffusion. As an example,  we take the sharp interface and numerically allow it to diffuse for a short fixed amount of time and then recalculate the resistance, as shown by the black solid curve 
  in Figure \ref{fig:bistable_exp_strat}a. 
  Our simple diffusion model recovers the fully mixed result in the limit of long diffusion times as shown by the red dash-dot curve 
  in Figure \ref{fig:bistable_exp_strat}a.

We repeat the experiments already presented by removing the mixers which were placed after each junction. We observe that the flow remains significantly stratified throughout the length of the tube. We again record the switching point in this stratified case where the flow spontaneously changes direction in branch C. This data is shown in Figure \ref{fig:bistable_exp_strat}b and is compared to the theoretical result assuming the fluids are fully mixed, completely stratified, and allowed to diffuse. A diffusion time of 0.05 agrees very well with our experimental result. Of course the true resistance law for miscible stratified flow would need to account for diffusion as a function of length down the tube, but our simple model seems to capture the essential behavior. Switching still occurs at $P_1/P_2=1$ whether the mixers are present or not, and we do not show this data here for clarity.

\section{Conclusions}

We have demonstrated bistability in a simple fluid experiment using two miscible fluids. Typical flow rates for our experiments are such that the maximum Reynolds number in any experiment was approximately 100 (though typically much less) and laminar flow within the network is expected. The only source of nonlinearity is the viscosity of the mixture which is enough to lead to spontaneous flow switching in the network. This prediction is robust and expected to occur in a wide range of fluid systems. Although we built a large scale network our predictions are scale free and equally apply to very small and very large diameter tubes as long as the basic assumption of laminar flow is maintained. Obviously more complicated networks with more nodes would be expected to display even more complicated behavior.

Based on the simplicity of the theory and the excellent agreement with experiments, we expect  that Eq. \ref{eq:existence_general} should serve as a general criteria for observing bistability in different applications. For example, we could extend our analysis and experiments to account for different effects such as turbulent flow, immiscible fluids, or concentrated suspensions.  We also note from Eq. \ref{eq:existence_general} that if we take the limit of the connecting branch C going to zero length, we can still obtain bistability. This limit is the case of two inlets and two outlets connected at a single node, which is similar to the microfluidic flip-flop studied by Groisman and Quake \cite{Groisman:2003p4579}. Their flip-flop required visco-elastic fluids whereas here we find bistability in a similar geometry with Newtonian fluids of different viscosities.

It is surprising to us that we can find no evidence that these predictions and experiments have previously been made or carried out. The experiments are simple to perform, the calculations involve elementary mathematics, and the physics has been well-established for over a century. This system also provides a striking example of bistability in a simple fluid network which, at first glance, does not seem to have the essential nonlinear ingredient.

The existence of bistability in this simple network also suggests that other networks could possibly be designed and constructed to provide a variety of logic and control devices. These preliminary experiments also don't rule out the possibility of observing dynamic behavior such as spontaneous oscillations. Finally, our results may also have relevance for microfluidic applications. In any device with even a simple network involving fluids of different viscosities or concentrated suspensions, we show here that we cannot predict the direction of flow in the network without knowing the flow's history.


\appendix
\section{Derivation of Bistability Criterion}

In the text we provide a simple criterion to determine whether a given network and fluid combination can exhibit bistability. In this section we provide a detailed derivation for the interested reader. The relationship between the flow in branch C to the inlet flows was given in the text
in Eq. \ref{eq:qc}. In what follows, it simplifies matters to use dimensionless flow rates where all flows are scaled by  the total flow, $Q_1+Q_2$. For simplicity of notation, for the remainder of this
Appendix, all flows are dimensionless though not annotated in any special way. Under this scaling Eq. \ref{eq:qc} becomes,
\begin{equation}
Q_C  = \frac{Q_1 R_A - (1-Q_1) R_{B}}{ R_A  + R_B + R_C},
\label{eq:normalized}
\end{equation}
where the dimensionless flow in C, $Q_C \in [-1,1]$, and the dimensionless flow in inlet 1, $Q_1 \in [0,1]$. Without loss of generality, we assume that the more viscous fluid always occupies inlet 2.

\begin{figure}
 \includegraphics[width=8cm]{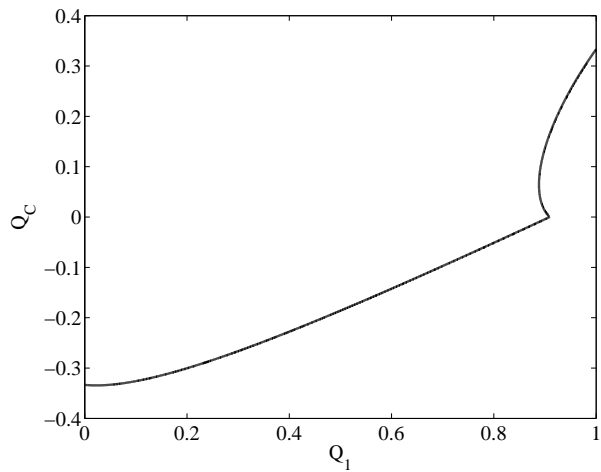}
\caption{Dimensionless flow in branch C, $Q_C$, versus dimensionless flow in branch 1, $Q_1$ for a viscosity
ratio of 10 and all nominal resistances are equal. }
\label{fig:qc_q1}
\end{figure}

An example of the equilibrium curve is shown in the $(Q_1,Q_C)$-plane in Figure \ref{fig:qc_q1}. If the more viscous solution occupies inlet 2 then it can be shown that $d Q_1/dQ_C$ is always positive as $Q_C$ approaches $0$ from below. A necessary and sufficient condition then for bistability is $d Q_1/dQ_C < 0$ as $Q_C$ approaches $0$ from above. Eq. (\ref{eq:normalized}) can be written as
\[ Q_C = \psi(Q_C,Q_1), \]
where the nonlinear function $\psi$ is defined as
\[ \psi = \frac{Q_1 R_A - (1-Q_1) R_{B}}{ R_A  + R_B + R_C}, \]
for the three-node network. The derivative of $Q_C$ with respect to $Q_1$ is
\[ \frac{dQ_C}{d Q_1} = \frac{\partial \psi}{\partial Q_1} + \sum_{j=A,B,C} \frac{\partial \psi}{\partial R_j} \frac{\partial R_j}{\partial \phi_j} \frac{d \phi_j}{d Q_1}. \]
For $Q_C>0$ the volume fraction in each of the branches is
\[ \phi_A = 0, \; \phi_C = 0, \; \phi_B = \frac{1-Q_1}{Q_C + (1 - Q_1)},\]
in which case $d Q_C/d Q_1$ reduces to
\begin{equation}
\frac{dQ_C}{d Q_1} = \frac{\partial \psi}{\partial Q_1} + \frac{\partial \psi}{\partial R_B} \frac{\partial R_B}{\partial \phi_B} \frac{d \phi_B}{d Q_1}.
\label{eq:qtheta}
\end{equation}
The derivatives are straight-forward to compute and are given by
\begin{eqnarray*}
\frac{\partial \psi}{\partial Q_1} &=& \frac{R_A + R_B}{R_A + R_B + R_C}, \\
\frac{\partial \psi}{\partial R_B} &=& \frac{-(R_A + (1-Q_1) R_C)}{(R_A + R_B + R_C)^2} = \frac{-(1 - (Q_1 -Q_C))}{R_A + R_B + R_C}, \\
\frac{d \phi_B}{d Q_1} &=& \frac{-Q_C - (1-Q_1) \frac{dQ_C}{dQ_1}}{(Q_C + (1-Q_1))^2}, \\
\frac{d R_B}{d \phi_B} &=& R_B \frac{d \ln(\mu_B)}{d \phi_B} .
\end{eqnarray*}
We can replace the derivatives into Eq. (\ref{eq:qtheta}) and solve for $d Q_1/dQ_C$.
In the limit as $Q_C \rightarrow 0^{+}$ we obtain,
\[ \frac{d Q_1}{d Q_C} = \frac{R_A + R_C + R_B (1 - \left. \frac{d \ln(\mu_B)}{d \phi_B} \right|_{\phi_B = 1})}{R_A + R_B}.\]
Since the denominator is always positive the condition $ d Q_1/dQ_C < 0$ for bistability becomes
\[ R_B \left( \left. \frac{d \ln(\mu_B)}{d \phi_B} \right|_{\phi_B = 1} - 1\right) > R_A + R_C. \]
Recall that as $Q_C \rightarrow 0$ from above $\mu_A = \mu_C = \mu_1$ and $\mu_B = \mu_2$. The expression can then be reduced to
\begin{equation}
\frac{\mu_2}{\mu_1} \left( \left. \frac{d \ln(\mu_B)}{d \phi_B} \right|_{\phi_B=1}- 1\right) > \frac{L_A D_B^4}{L_B D_A^4} + \frac{L_C D_B^4}{L_B D_C^4},
\label{eq:general}
\end{equation}
which is the general result quoted in the text.

The specific result quoted in the text for a fluid that obeys an Arrenhius law can be derived as follows. The Arrenhius law can be expressed as
\[ \ln(\mu) =   \ln(\mu_1) +  N \ln(\frac{\mu_2}{\mu_1}),\]
where $N$ is the mole fraction of fluid 2. The derivative we are looking for is then
\begin{equation}
\left. \frac{d \ln(\mu)}{d \phi} \right|_{\phi = 1} = \ln(\frac{\mu_2}{\mu_1}) \left. \frac{d N}{d \phi} \right|_{\phi = 1}.
\label{eq:muderiv}
\end{equation}
The relationship between volume fraction, $\phi$, and mole fraction, $N$, is given as
\[ N = \frac{ \phi \frac{\rho_2}{M_2}}{\phi \frac{\rho_2}{M_2} + (1-\phi) \frac{\rho_1}{M_1}}
= \frac{ 1}{ 1 + \left(1-\frac{1}{\phi} \right) \frac{\rho_1}{M_1} \frac{M_2}{\rho_2}},
\]
which implies that the derivative is
\begin{equation}
\left. \frac{d N}{d \phi} \right|_{\phi = 1} = \frac{\rho_1 M_2}{\rho_2 M_1}.
\label{eq:Nderiv}
\end{equation}
Replacing Eqs. (\ref{eq:muderiv}) and (\ref{eq:Nderiv}) into Eq. (\ref{eq:general}) recovers the specific result quoted in the text
\begin{equation}
\frac{\mu_2}{\mu_1} \left( \ln\left( \frac{\mu_2}{\mu_1} \right) \frac{\rho_1 M_2}{\rho_2 M_1} - 1\right)  >
\frac{L_A D_B^4}{L_B D_A^4}+\frac{L_C D_B^4}{L_B D_C^4}.
\end{equation}

\end{document}